\documentclass[11pt]{llncs}
\usepackage{fullpage}

\newcommand{\Oh}[1]
    {\ensuremath{\mathcal{O}\!\left({#1}\right)}}
\newcommand{\rank}
    {\ensuremath{\mathsf{rank}}}

\begin{document}

\title{Sequential-Access FM-Indexes}
\author{Travis Gagie}
\institute{Department of Computer Science and Engineering\\Aalto University, Finland}

\maketitle

\begin{abstract}
Previous authors have shown how to build FM-indexes efficiently in external memory, but querying them efficiently remains an open problem.  Searching na\"{i}vely for a pattern $P$ requires \(\Theta (|P|)\) random access.  In this paper we show how, by storing a few small auxiliary tables, we can access data only in the order in which they appear on disk, which should be faster.
\end{abstract}

An FM-index~\cite{FM05} is a compressed representation of a text that allows us to quickly search for arbitrary patterns in that text.  Their growing popularity in genomics (e.g., in BWT-SW, Bowtie, SOAP2 and BWA) means we should look for ways in which they can handle massive datasets, which may have to reside in external memory even when compressed.  Unfortunately, although we know how to build FM-indexes efficiently in external memory~\cite{FGM12}, querying them efficiently remains an open problem.  Searching na\"{i}vely for a pattern $P$ requires \(\Theta (|P|)\) random access, which are expensive due to seek times.  We refer the reader to the papers by Chien et al.~\cite{CHSV08}, Hon et al.~\cite{HSV10} and Ferragina~\cite{Fer10} for more discussion of this problem.  In this paper we extend a result by Orlandi and Venturini~\cite{OV11} to show how, by storing a few small auxiliary tables, we can access data only in the order in which they appear on disk.  We may read slightly more data but, since sequential access to disk is orders of magnitude faster than random access, our modified index should be faster overall.

FM-indexes are based on the Burrows-Wheeler Transform (BWT), which permutes the characters of a string $T$ based on the contexts that follow them.  We can compute \(B = \mathrm{BWT} (T)\) by lexicographically sorting the rotations of $T$, then recording the last character of each rotation.  (If we want to recover $T$ later, we append a special symbol before computing $B$ or record the position to which a designated character is mapped.)  For example, if
\[T = 110111100101110101010001111\,,\]
then
\[B = 110111011001001011111010110\,.\]
We use binary strings for simplicity but the results in this paper extend to any reasonable alphabet size.  Notice that, for any pattern $P$, the characters immediately preceding occurrences of $P$ in $T$ are adjacent in $B$ (considering $T$ to be cyclic). For example, if \(P = 0101\) then the characters immediately preceding occurrences of $T$ are \(T [8], T[14]\) and \(T [16]\), which are mapped to \(B [7], B [6]\) and \(B [5]\), respectively.  We call \(B [5..7]\) the interval for \(P = 0101\).

The basic operation of FM-indexes is to find the interval in $B$ for any given pattern $P$.  For example, the length of the interval is the number of occurrences of $P$ in $T$.  Notice that the left endpoint of the interval is the rank of the lexicographically first rotation of $T$ that starts with $P$, and the right endpoint is the rank of the lexicographically last such rotation.  To find these endpoints, we store data structures such that, for any character $c$ in the alphabet and any position $i$ in $B$, we can quickly compute the number \(\rank_c (i)\) of occurrences of $c$ in \(B [1..i]\).  We also store the number \(C [c]\) of characters in $B$ lexicographically less than $c$.

Suppose we are na\"{i}vely searching for the right endpoint of the interval; finding the left endpoint is essentially symmetric.  We iteratively compute
\begin{eqnarray*}
j_1 & = & \rank_{P [|P|]} (|B|) + C \left[ \rule{0ex}{2ex} P [|P|] \right]\,,\\[1ex]
j_2 & = & \rank_{P [|P| - 1]} (j_1) + C \left[ \rule{0ex}{2ex} P [|P| - 1] \right]\,,\\[1ex]
j_3 & = & \rank_{P [|P| - 2]} (j_2) + C \left[ \rule{0ex}{2ex} P [|P| - 2] \right]\,,\\
& \vdots &\\
j_{|P|} & = & \rank_{P [1]} (j_{|P| - 1}) + C [P [1]]\,;
\end{eqnarray*}
by induction, $j_{|P|}$ is the right endpoint.  In our example \(C = [0, 10]\), so we compute
\begin{eqnarray*}
j_1 & = & \rank_1 (27) + C [1] = 27\,,\\[1ex]
j_2 & = & \rank_0 (27) + C [0] = 10\,,\\[1ex]
j_3 & = & \rank_1 (10) + C [1] = 17\,,\\[1ex]
j_4 & = & \rank_0 (17) + C [0] = 7\,.
\end{eqnarray*}
Unfortunately, with this method, the sequence of positions for which we answer \rank\ queries can be far from ordered and so, with current data structures supporting those queries, we use many random access.

Our idea is to build a series of small auxiliary tables that appear on disk before $B$, in column-major order.  Each of these tables stores the answers to $\rank_c$ queries for each character $c$, sampled at evenly spaced positions.  The sample rate increases geometrically from each table to the next.  For our example we might store two tables in addition to $B$,
\[\begin{array}{l@{\hspace{2ex}}l}
\rank_0 (9) = 2 & \rank_1 (9) = 7\\
\rank_0 (18) = 7 & \rank_1 (18) = 11\\
\rank_0 (27) = 10 & \rank_1 (27) = 17\\[5ex]
\rank_0 (3) = 1 & \rank_1 (3) = 2\\
\rank_0 (6) = 1 & \rank_1 (6) = 5\\
\rank_0 (9) = 2 & \rank_1 (9) = 7\\
\rank_0 (12) = 4 & \rank_1 (12) = 8\\
\rank_0 (15) = 6 & \rank_1 (15) = 9\\
\rank_0 (18) = 7 & \rank_1 (18) = 11\\
\rank_0 (21) = 7 & \rank_1 (21) = 14\\
\rank_0 (24) = 9 & \rank_1 (24) = 15\\
\rank_0 (27) = 10 & \rank_1 (27) = 17
\end{array}\]

Assume our first table is small enough to fit into main memory.  We need no \rank\ queries to compute \(j_1 = 27\); nor do we need any to compute \(j_2 = 10\), although that is just because \(P [4]\) is an occurrence of the largest character in the alphabet.  To compute $j_3$ exactly we need \(\rank_1 (10)\), which we do not have stored.  However, we estimate \(\rank_1 (10) \approx \rank_1 (18) - (18 - 10) = 3\), which leads to the estimate \(j_3 \approx 13\).  We then estimate \(\rank_0 (13) \approx \rank_0 (18) - (18 - 13) = 5\), which leads to the estimate \(j_4 \approx 5\).

Orlandi and Venturini~\cite{OV11} pointed out that, if \(j - \ell \leq i \leq j\), then \(\rank_c (j) - \min (j - i, \ell) \leq \rank_c (i) \leq \rank_c (j)\), which means that our estimate \(j_3 \approx 13\) is a lower bound within 8 of the true value; in general, our error can be as large as the distance between samples, which is 9 in this case.  The surprising part of their result is that our error cannot exceed this distance, even after repeated estimations using this formula.  Therefore, our estimate \(j_4 \approx 5\) is also a lower bound within 9 of the true value.

We now discard the first table and consider what information we want from the second table.  We know \(j_1 = 27\), \(j_2 = 10\), \(13 \leq j_3 \leq 22\) and \(5 \leq j_4 \leq 14\), considering always only the loose error bound 9; we want to re-estimate \(\rank_1 (10)\) and \(\rank_0 (j_3)\) in order to re-estimate $j_3$ and $j_4$.  Therefore, we want to read the values \(\rank_1 (12) = 8\) to re-estimate $j_3$; depending on that re-estimate of $j_3$, we will use one of the values \(\rank_0 (15) = 6\), \(\rank_0 (18) = 7\), \(\rank_0 (21) = 7\) and \(\rank_0 (24) = 9\) to re-estimate $j_4$.  We consider where all these values appear in the second table (notice they form two consecutive blocks; generally there will be one block for each value we are trying to estimate), sort the positions, and then read them sequentially.

We re-estimate \(\rank_1 (10) \approx \rank_1 (12) - (12 - 10) = 6\), which leads to the re-estimate \(j_3 \approx 16\).  We then estimate \(\rank_0 (16) \approx \rank_0 (18) - (18 - 16) = 5\), which leads to the re-estimate \(j_4 = 5\).  Our re-estimates of $j_3$ and $j_4$ are again lower bounds, this time within 3 of the true values.  We now discard the data we have read from the second table and consider what data we want to read from $B$ itself, sort the positions, and then read them sequentially.  Details of how we do this depend on which data structures we use to support \rank\ queries on $B$, but now the sequence of positions for which we answer \rank\ queries is ordered.

Calculation shows that, if we increase the sample rate by a factor of $r$ between each table, then we use $\Oh{\log_r |T|}$ tables of total size $\Oh{\sigma |T| \log |T| / r}$ bits, where $\sigma$ is the size of the alphabet.  For reasonable values of $r$ and $\sigma$, this space bound should usually be small compared to $B$ itself, and may be reducible with clever encoding of the tables.  To find the interval for $P$, we read a total of $\Oh{r |P| \log^2 |T| / \log r}$ bits from the tables, which is more than what we would read with the na\"{i}ve method, but we read them sequentially.

\bibliographystyle{plain}
\bibliography{seq_acc_fm}

\end{document}